\documentclass{eas}
\usepackage[latin1]{inputenc}
\usepackage[comma,authoryear]{natbib}
\usepackage[dvips]{graphicx}

\def\bbbc{{\mathchoice {\setbox0=\hbox{$\displaystyle\rm C$}\hbox{\hbox
to0pt{\kern0.4\wd0\vrule height0.9\ht0\hss}\box0}}
{\setbox0=\hbox{$\textstyle\rm C$}\hbox{\hbox
to0pt{\kern0.4\wd0\vrule height0.9\ht0\hss}\box0}}
{\setbox0=\hbox{$\scriptstyle\rm C$}\hbox{\hbox
to0pt{\kern0.4\wd0\vrule height0.9\ht0\hss}\box0}}
{\setbox0=\hbox{$\scriptscriptstyle\rm C$}\hbox{\hbox
to0pt{\kern0.4\wd0\vrule height0.9\ht0\hss}\box0}}}}
\def\bbbq{{\mathchoice {\setbox0=\hbox{$\displaystyle\rm
Q$}\hbox{\raise
0.15\ht0\hbox to0pt{\kern0.4\wd0\vrule height0.8\ht0\hss}\box0}}
{\setbox0=\hbox{$\textstyle\rm Q$}\hbox{\raise
0.15\ht0\hbox to0pt{\kern0.4\wd0\vrule height0.8\ht0\hss}\box0}}
{\setbox0=\hbox{$\scriptstyle\rm Q$}\hbox{\raise
0.15\ht0\hbox to0pt{\kern0.4\wd0\vrule height0.7\ht0\hss}\box0}}
{\setbox0=\hbox{$\scriptscriptstyle\rm Q$}\hbox{\raise
0.15\ht0\hbox to0pt{\kern0.4\wd0\vrule height0.7\ht0\hss}\box0}}}}
\def\bbbt{{\mathchoice {\setbox0=\hbox{$\displaystyle\rm
T$}\hbox{\hbox to0pt{\kern0.3\wd0\vrule height0.9\ht0\hss}\box0}}
{\setbox0=\hbox{$\textstyle\rm T$}\hbox{\hbox
to0pt{\kern0.3\wd0\vrule height0.9\ht0\hss}\box0}}
{\setbox0=\hbox{$\scriptstyle\rm T$}\hbox{\hbox
to0pt{\kern0.3\wd0\vrule height0.9\ht0\hss}\box0}}
{\setbox0=\hbox{$\scriptscriptstyle\rm T$}\hbox{\hbox
to0pt{\kern0.3\wd0\vrule height0.9\ht0\hss}\box0}}}}
\def\bbbz{{\mathchoice {\hbox{$\sf\textstyle Z\kern-0.4em Z$}}
{\hbox{$\sf\textstyle Z\kern-0.4em Z$}}
{\hbox{$\sf\scriptstyle Z\kern-0.3em Z$}}
{\hbox{$\sf\scriptscriptstyle Z\kern-0.2em Z$}}}}
\newcommand{\qqqquad}{\qquad\qquad}

\newcommand{\BV}{Brunt-V\"ais\"al\"a\ }

\newcommand{\beq}{\begin{equation}}
\newcommand{\beqa}{\begin{eqnarray*}}
\newcommand{\beqan}{\begin{eqnarray}}
\newcommand{\greq}{\begin{equation}\left\{ \begin{array}{l}}
\newcommand{\egreq}{\end{array}\right. \end{equation}}
\newcommand{\nngreq}{\[\left\{ \begin{array}{l}}
\newcommand{\nnegreq}{\end{array}\right. \]}
\newcommand{\egreqn}[1]{\end{array}\right. \label{#1}\end{equation}}
\newcommand{\eeq}{\end{equation}} 
\newcommand{\eeqn}[1]{\label{#1}\end{equation}} 
\newcommand{\eeqa}{\end{eqnarray*}}
\newcommand{\eeqan}[1]{\label{#1}\end{eqnarray}}

\newcommand{\ssi}{ \Longleftrightarrow}

\newcommand{\lp}{ \left(}
\newcommand{\rp}{ \right)}
\newcommand{\lc}{ \left[}
\newcommand{\rc}{ \right]}
\newcommand{\la}{ \left\{}
\newcommand{\ra}{ \right\}}
\newcommand{\khi}{\chi}

\newcommand{\ulm}{u^\ell_m}
\newcommand{\vlm}{v^\ell_m}
\newcommand{\wlm}{w^\ell_m}

\newcommand{\YL}{Y^m_\ell}

\newcommand{\RL}{ \vec{R}^m_\ell }
\newcommand{\SL}{ \vec{S}^m_\ell }

\newcommand{\TL}{ \vec{T}^m_\ell }

\newcommand{\eps}{\varepsilon}

\newcommand{\na}{ \vec{\nabla} }

\newcommand{\cth}{ \cos\theta }

\newcommand{\sth}{ \sin\theta }

\newcommand{\RA}{\mbox{$ {\rm Ra} $}}

\newcommand{\vu}{\vec{u}}

\newcommand{\er}{\vec{e}_r}

\newcommand{\ephi}{\vec{e}_\varphi}

\newcommand{\es}{\vec{e}_s}
\newcommand{\ex}{\vec{e}_x}
\newcommand{\ey}{\vec{e}_y}
\newcommand{\ez}{\vec{e}_z}

\renewcommand{\vg}{\vec{g}}

\newcommand{\vs}{\vec{s}}

\newcommand{\vv}{\vec{v}}

\newcommand{\vo}{\vec{\omega}}
\newcommand{\vO}{\vec{\Omega}}

\newcommand{\vzero}{\vec{0}}

\newcommand{\od}[1]{\mbox{${\cal O}(#1)$}}

\newcommand{\dr}[1]{\frac{\partial  #1}{\partial r}}

\newcommand{\dt}[1]{\frac{\partial  #1}{\partial t}}

\newcommand{\dnr}[1]{\frac{d  #1}{dr}}

\newcommand{\dnz}[1]{\frac{d  #1}{dz}}

\newcommand{\Dt}[1]{\frac{D #1}{D t}}

\newcommand{\dz}[1]{\frac{\partial  #1}{\partial z}}

\newcommand{\ddx}[1]{\frac{\partial^2  #1}{\partial x^2}}

\newcommand{\ddz}[1]{\frac{\partial^2  #1}{\partial z^2}}

\newcommand{\drsint}[1]{\frac{1}{r\sin\theta}\frac{\partial}{\partial\theta}(\sin\theta #1)}

\newcommand{\drr}{\frac{\partial}{\partial r}}

\def\Div{\mathop{\hbox{div}}\nolimits}

\def\rot{\mathop{\hbox{\overrightarrow{\displaystyle\rm Rot}}}\nolimits}

\newcommand{\disp}[1]{\displaystyle #1}
\newcommand{\eq}[1]{(\ref{#1})}

\begin{document}
\title{On the dynamics of radiative zones in rotating stars}
\author{Michel Rieutord}
\address{
Laboratoire d'Astrophysique de l'observatoire Midi-Pyrénées, UMR 5572,
CNRS et Université Paul Sabatier, 14 avenue E. Belin, 31400 Toulouse,
France}

\date{\today}
\bibliographystyle{apj}

\begin{abstract}
In this lecture I try to explain the basic dynamical processes at
work in a radiative zone of a rotating star. In particular, the notion
of baroclinicity is thoroughly discussed. Attention is specially
directed to the case of circulations and the key role of angular momentum
conservation is stressed. The specific part played by viscosity is
also explained. The old approach of Eddington and Sweet is reviewed and
criticized in the light of the seminal papers of Busse 1981 and Zahn 1992.
Other examples taken in the recent literature are also presented;
finally, I summarize the important points.
\end{abstract}

\maketitle

\section{Introduction}

It is well known since the seminal work of \cite{VZ24} that radiative
zones of a rotating star cannot be at rest; in other words no
hydrostatic solution exists in any frame: some shearing flow must occur.
The importance of such flows comes from their relation with the
abundances of elements at the surface of stars. Indeed, the
interpretation of these observations requires the understanding of
the close relation between rotation and mixing, namely the transport of
elements in radiative zones, which, without rotation would be a pure
diffusive process.

As any fluid dynamical process, flows in rotating radiative zones are
governed by PDE and therefore difficult to evaluate in a simple way.
This is the reason why much misleading work has been published on the
subject, especially under the title of the Eddington-Sweet circulation.
In this lecture I would like to present as clearly as possible the
dynamical processes at work in a rotating radiative zone and explain why the
idea of Eddington-Sweet circulation is a misunderstanding of the
problem.

\section{Dynamical ingredients}

\subsection{Equations of motion}

Unlike many  problems in fluid mechanics where flows are produced by a
given force field (just think for instance to the case of a river whose
water flows down thanks to gravity), flows in rotating radiative
zones are the result of a mismatch between two force fields, namely the
pressure gradient and the gravity force. This mismatch gives birth to a
torque field, the baroclinic torque, which locally produces vorticity and
thus fluid motion. Before discussing this matter, it is however
necessary to have a look to the equations of motion and especially to
the momentum equation in a rotating frame; this equation reads:

\[ \rho\lp \Dt{\vv}+ \underbrace{2\vO\wedge\vv}_{\rm I}+
\underbrace{\Omega^2\vs}_{\rm II}\rp = -\na P +
\underbrace{\rho\vg}_{\rm III}\]
where we discarded viscosity. Terms I, II, III are specific of this
problem; they are respectively for:

\begin{itemize}
\item The Coriolis acceleration $2\vO\wedge\rho\vv$ which insures
conservation of angular momentum.
\item The centrifugal acceleration $-\Omega^2\vs$ which is at the origin
of baroclinicity.
\item The gravity force which transform a density fluctuation into
motion.
\end{itemize}

\subsection{The wave zoo}

Before diving into the heart of the subject, it is useful to recall the
waves able to propagate in a rotating radiative zone. There are three
sorts of waves of that kind: the acoustic, the gravity and the inertial
waves.

\subsubsection{The acoustic waves}

In a uniform medium acoustic waves are described by

\greq
\rho_0\dt{\vv} = -\na p'\\
\dt{\rho'} + \rho_0\Div\vv = 0 \\
\dt{s'}=0
\egreq
namely momentum, mass and entropy conservation. Note that for such waves
$\partial_t\rho'$ is an essential term. Equations have been linearized
and 0-index terms refer to the background state. The dispersion relation
of these waves, namely $\omega=c_sk$, where $c_s$ is the speed of sound,
shows that acoustic frequencies have no upper limit. But they have a
lower limit given by a $k_{\rm min}$ coming from the finite size of the
container (the star !).

\subsubsection{The gravity waves}

While acoustic waves exist because of the pressure field, the restoring
force of gravity waves is the buoyancy. Compressibility is unnecessary
and therefore density can be kept constant except when it yields buoyancy
(this is the Boussinesq approximation in a very short description). In
its simplest form, the system of equation describing these waves is

\greq
\rho_0\dt{\vv} = -\na p'+\rho'\vg\\
\Div\vv=0\\
\dt{T'} + \vv\cdot\na T_0 = 0\\
\rho'/\rho_0 = -\alpha T'
\egreq
where $\alpha$ is the dilation coefficient. Note that temperature,
which could be replaced by any other scalar field (like concentration
for instance), needs to be non-uniform.

Unlike acoustic waves, gravity waves suffer an anisotropy imposed by
the direction of gravity and temperature gradient. This results in a
dispersion relation of the form:

\[ \omega = N\frac{\sqrt{k_x^2+k_y^2}}{k}\]
where gravity and $\na T_0$ are along $\ez$; $N=\lp -\alpha g
\dnz{T_0}\rp^{1/2}$ is the \BV\ frequency. This
relation shows that the wave frequency is bounded by $N$ and therefore
gravity waves occupy the low-frequency band.

\subsubsection{The inertial waves}

Lastly, waves induced by the Coriolis force. Inertial waves are the
solution of

\greq
\rho_0\lp\dt{\vv}+2\vO\wedge\vv\rp = -\na p'\\
\Div\vv=0
\egreq
and thus contain only a velocity and a pressure perturbation. Like the
gravity wave they behave anisotropically. The direction of the rotation
axis gives the preferred direction. Plane waves obey the following
dispersion relation:

\[ \omega = 2\Omega\frac{k_z}{k}\]
which shows that these waves are also in the low-frequency range,
bounded by $2\Omega$ the Coriolis frequency \cite[see][]{Green69,
rieutord97}.

\subsubsection{Some more waves}

One often meets, especially in the geophysical literature, other waves
like Rossby, baroclinic... waves. These are subclasses of the foregoing
(gravity and inertial) waves meeting some additional constrains or
evolve on some specific flows. For instance, Rossby waves are inertial
waves which do not (or little) depend on the coordinate parallel to the
rotation axis, while baroclinic waves combine gravity and coriolis forces
and surf on shear flows.

\subsubsection{Further remarks}

As inertial waves and gravity waves occupy the low-frequency range, they
can perturb each other very strongly. The common situation in stars is
that \mbox{$N>2\Omega$}. This means that gravity modes of frequency near
$2\Omega$ or lower, are strongly perturbed. In that case, gravity and
inertial modes are undistinguishable and one speaks about
gravito-inertial waves \cite[see][]{DRV99}. 

When dealing with rotation we did not mention the centrifugal acceleration;
this is because this field does not generate any specific wave. It
modifies an existant gravity field and therefore just enters the
buoyancy force.

\section{Baroclinicity}

\subsection{Introduction to baroclinicity}\label{sect_th}

Baroclinicity means the inclination of isobars with respect to
``horizontality" usually given by equipotentials. More generally, if
isotherms, isodensities, isobars and equipotentials are not identical,
the fluid is said in a baroclinic state while if isodensities and isobars
(and thus isotherms) are identical the fluid is said barotropic. The
importance of this distinction comes from the existence or inexistence
of hydrostatic solutions. Let us examine the necessary condition for a
fluid to be in hydrostatic equilibrium:

\beqa
-\na P -\rho\na\Phi &=& \vzero \quad{\rm Mechanical\; equilibrium} \\
\Div(\khi\na T) &=& 0\quad{\rm Thermal\; equilibrium} \\
f(P,\rho,T)&=&0 \quad{\rm Equation\; of \; state}
\eeqa

$\na\Phi$ is either prescribed or obtained from $\Delta\Phi=4\pi
G\rho-2\Omega^2$. In order that mechanical equilibrium be possible, it
is necessary that:

\[ \rot\lp\frac{1}{\rho}\na P\rp = \vzero \ssi \na\rho\wedge\na P
=\vzero\]
i.e. that isobars and isodensities are identical surfaces. Now we may
differentiate the equation of state,

\[ \frac{\partial f}{\partial P}dP+\frac{\partial f}{\partial\rho}d\rho
+ \frac{\partial f}{\partial T}dT = 0\]
Considering that mechanical equilibrium is realized, on an isobar (which
is an isodensity), we have

\[ \frac{\partial f}{\partial T}dT = 0\]
which shows that either the isobar is an isotherm, $dT=0$, or $\frac{\partial
f}{\partial T} = 0$ i.e. the fluid is barotropic and there is a relation
between pressure and density, $P\equiv P(\rho)$. This latter case is
rather specific and usually found because either the fluid is considered
to be isentropic or isothermal. In general, $\frac{\partial f}{\partial T}
\neq 0$ and isotherms, which are determined by the equation of thermal
equilibrium, are not identical with isobars. This general case is
illustrated on figure 1.

\begin{figure}
\centerline{\includegraphics[width=6cm,angle=0]{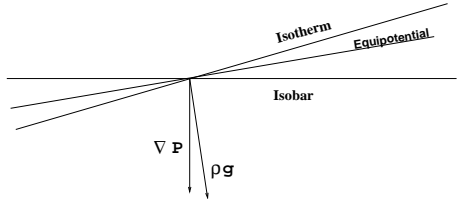}}
\caption[]{Inclination of an isobar, an isotherm and an equipotential in
a baroclinic set-up.}
\end{figure}

\subsection{Examples}

To illustrate the notion of baroclinicity we shall consider two
classical examples which can be solved explicitely.

\subsubsection{Double glazing window}

The first example comes from the classical problem of natural convection
which arises within a fluid imprisoned between two vertical plates at
different temperatures; such a situation is the one prevailing inside
isolating windows with double glazing where air is trapped between two
glass panes, a cold and a warm one.

Since temperature fluctuations are small, we assume the Boussinesq
approximation and neglect all density fluctuations except those
generating the buoyancy. Such a small amplitude motion is governed by:

\greq
\disp{-\na \delta P +\delta\rho \vg + \mu \Delta\vv}=0 \\
\disp{\vv\cdot\na T_{eq} = \kappa\Delta\delta T} \\
\Div\vv=0\\
\disp{\frac{\delta\rho}{\rho}=-\alpha\delta T}
\egreqn{bbc}
Recall that in this problem $\na T_{eq}=\vzero$ since at equilibrium
temperature needs to be constant.

\begin{figure}
\centerline{\includegraphics[width=8cm,angle=0]{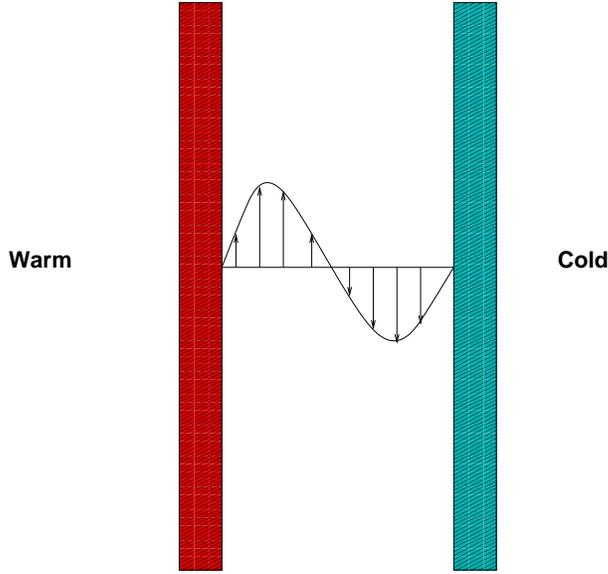}}
\caption{Flow between two vertical plates at different temperatures.
This picture illustrates the shear flow generated by the baroclinic
torque.}
\label{double_vitre}
\end{figure}

Let us assume that the fluid is between two plates, infinite in the $y$
and $z$ directions separated by $d$ in horizontal $x$-direction, with
the warm plate at $T=T_c$ (resp. cold at $T=T_f$) in $x=-d/2$ (resp. in
$x=d/2$). Thermal equilibrium leads to

\[ \delta T(x) = - (T_c-T_f) x /d \]
\[ \delta\rho = \frac{\alpha(T_c-T_f)\rho_0}{d}x \]
and thus the velocity field verifies:

\[ \mu \Delta\vv -\na \delta P = \frac{\alpha(T_c-T_f)\rho_0g}{d}x\ez,
\qquad \Div\vv=0\]
in this problem the baroclinic torque is balance by the viscous torque:

\[ \mu\Delta\vo = \frac{\alpha(T_c-T_f)\rho_0g}{d}\ex\wedge\ez, \qquad
\vo = \na\times\vv \]
The solution of this equation is easily derived and may be expressed as:

\beq v_z(x) = \frac{\alpha(T_c-T_f)\rho_0g}{24\mu d} x (d^2-4x^2) \eeq
where we used no-slip boundary conditions on the plates together with a
zero-vertical mass flux to determine the three constants of integration.
The flow thus obtained is shown in figure~\ref{double_vitre}.

This is a vortical shear flow where vorticity is generated by the torque
and diffused to the walls by viscosity.

\subsubsection{Thermal wind}

The second example is more closely related to our astrophysical concern.
It is issued from geophysics and describes the baroclinic situation that
occurs in the Earth's atmosphere where solar heating imposes a
latitudinal temperature gradient; poles which are cooler than the
equator may be compared to the cold pane of the foregoing example. If
the Earth were not rotating the baroclinic torque would induce, like
above, a circulation between poles and equator. However, Earth rotates
and therefore air cannot move easily from pole to equator because of
angular momentum conservation. In this case the baroclinic torque
induces an azimuthal flow; the mechanism is just similar to the one of a
precessing spinner: the torque induced by the gravitationnal field on
the spinner induces an azimuthal motion, namely the precession.

Let's have a look to the equation of the flow using the Boussinesq
approximation; the steady linearized inviscid momentum equation reads:

\[ 2\vO\wedge\vv = -\frac{1}{\rho_0}\na\delta P -\alpha\delta T\vg\]
where we neglected viscous force and non-linear terms. Here the
baroclinic torque is balanced by the torque issued from Coriolis
acceleration, namely

\[ \rot(2\vO\wedge\vv) = \rot(\alpha g\delta T\ez)\]
More explicitely, assuming that $\delta T\equiv\delta T(x)$ ($x$ being
the North-South direction), we find 

\[\dz{v_y} \propto \delta T'(x) \]
This equation shows the appearance of an azimuthal $v_y$ flow and an
axial gradient of angular momentum.

Anticipating the next section, we see that the first effect of a
(baroclinic) torque is, as in the precessing spinner, to change the
angular momentum distribution and not to induce a meridian
circulation.

\section{Rotating radiative zones - 1924 -- 2004 : the evolution of
ideas in eighty years}

\subsection{von Zeipel theorem 1924}

In 1924 von Zeipel addressed the question of the equilibrium of a rotating
radiative zone.  Basically, he considered the
point of section \S\ref{sect_th} in the stellar context. Let us
remind us his result. If we assume that a radiative rotating envelope is
in hydrostatic equilibrium, then, as we showed in \ref{sect_th},
isobars, isochores, isotherms are identical and therefore all
thermodynamical quantities should depend only on, say, the total
potential $\Phi$ (centrifugal and gravitational); if the thermal equilibrium
is realized 

\[\Div(\khi\na T) + \rho\eps = 0 \qquad {\rm or} \qquad
\Div(\khi(\Phi)\na T(\Phi)) +\rho\eps= 0  \]

\[ \ssi\qquad \khi T'\Delta\Phi + (\khi T')'(\na\Phi)^2 +\rho\eps =0\]
where $\eps$ is the rate of energy generation per unit mass. Using
Poisson equation for the effective potential,

\beqa \underbrace{\khi(\Phi) T'(\Phi)(4\pi G\rho(\Phi)-2\Omega^2)}_{\rm
Cte} +
 \underbrace{(\khi(\Phi)T'(\Phi))'}_{\rm
Cte}\underbrace{(\na\Phi)^2}_{\rm not\; Cte} +\rho\eps =0\eeqa
where $\na\Phi$ is just the effective gravity which is not constant on
an equipotential (\mbox{$\vg_{\rm pole}\neq \vg_{\rm eq}$}).
With such an equation, von Zeipel correctly concluded that an equilibrium
solution is possible only if $(\khi(\Phi)T'(\Phi))'=0$ and if

\[ \eps = C\lp1-\frac{\Omega^2}{2\pi G\rho}\rp\]
Apparently, von Zeipel was pleased with such a result. But
\cite{Eddington1925} was not, being skeptical that energy generation,
presumably of microscopic origin, could follow such a law.
He suggested\footnote{Independently, Vogt 1925 had the same idea.} that
hydrostatic equilibrium should be abandoned and that a
meridian circulation would result from the thermal imbalance, putting
forward the picture that hot poles and cool equator (or vice versa)
would lead to some global convection with rising material at poles and
sinking one at equator (or vice versa).

Later, the solution of von Zeipel was qualified as paradoxical by
\cite{opik51} because of being not physically sound. I suppose that it is
since that time that von Zeipel result is also quoted as ``von Zeipel
paradox''. Presently however, von Zeipel paradox rather refers to the
inexistence of a static radiative equilibrium in a uniformly rotating
star \cite[]{busse82,HK94}.

\subsection{The Eddington-Sweet answer to von Zeipel paradox}

The ideas of Eddington and Vogt have been transformed into a model by
Sweet in 1950. The reasoning is the following: since $\Div(\khi\na T)\neq
0$ and since we are looking for a steady solution, the velocity
field which appears must be such that:

\[ \rho c_v(\vv\cdot\na T) = \Div(\khi\na T) \]
Assuming the thermal imbalance, $\Div(\khi\na T)$, is given, the
temperature field being almost spherically symmetric, this equation
gives an expression of the radial velocity:

\[ \rho c_v v_r\dr{T} \simeq  \Div(\khi\na T)\]
while $v_\theta$ is obtained from mass conservation:

\[ \Div(\rho\vv) =  \frac{1}{r^2}\dr{}(r^2\rho v_r)+\drsint{\rho
v_\theta}\]
The problem is that such a velocity field is not produced by a force field
and therefore the conservation of angular momentum is not insured since
it is not a solution of the momentum equation.

Such a difficulty was already pointed out by \cite{Randers41} who
underlined the key role of viscosity and questioned the existence of
meridian currents in an inviscid star. Then, \cite{Schwarz47,BK59} and
\cite{roxburgh64} showed more and more explicitly that some differential
rotation permits the existence of (inviscid) rotating radiative zone
without meridian circulation\footnote{Actually, Milne, quoted by
\cite{Eddington1925}, also proposed such a possibility.}. These authors
actually pointed out the existence of the thermal wind solutions.

\cite{KW90} in {\em Stellar structure and evolution} note
the problem of angular momentum conservation and the existence of
solutions without meridian circulation but do not show the link
between the two properties.

In fact , the connection between differential rotation, viscosity and
meridian circulation were explained by Busse \cite[]{busse81,busse82}
and the first self-consistent solutions (up to a turbulence model) were
given in a seminal paper by \cite{zahn92}.

We need now to review these two landmarks of the subject.

\subsection{Busse 1981 : ``Do Eddington-Sweet circulation exist?''}

The problem faced by von Zeipel was presented in terms of a thermal
imbalance which, through buoyancy (hot poles and cold equator), would
drive a meridian circulation. However, such a steady motion is
impossible in a rotating fluid because the angular momentum is not
locally conserved: the flux of angular momentum induced by the
meridian circulation need to be compensated by another flux (through
magnetic or viscous forces for instance).

Such steady motion are thus impeded mechanically if we ignore all other
forces than buoyancy. Another possibility is of course to relax the
steadiness hypothesis; however, in such a case conservation of angular
momentum appears as a restoring force (namely Coriolis force) which will
transform any initial perturbation into a (inertial) wave motion whose
transport properties are another story\footnote{Waves, although being an
essentially periodic motion, can transport momentum or chemicals
through their mean motion when they are of finite amplitude. }.

However, if the situation is stuck mechanically, it can evolve
thermally. This is what occurs indeed. In this way the fluid may move...
This remark underlines the real nature of the problem as it has be
enounced: this is an initial value problem; then, if the time scale of
the transient is short enough the star reaches a steady state or, if it is
too long, initial conditions should be of dramatic influence.

To illustrate the foregoing discussion, we shall follow the work of
\cite{busse81} and his example in cartesian geometry. With such a
set-up, one can consider arbitrary initial conditions and compute the
flow and its time scales in every detail.

Busse considered a fluid with very small Prandtl number (as expected in
a radiative stellar envelope) so that viscosity can be neglected in the
first instance. Thus, the only source of (long term) evolution is
thermal diffusion which damps out the transients to the steady state.
According to our foregoing remarks which pointed the impossibility of
steady meridian circulation without viscosity or magnetic field, we
expect that such meridian flow occurs only during the transient.

\begin{figure}
\centerline{\includegraphics[width=10cm,angle=0]{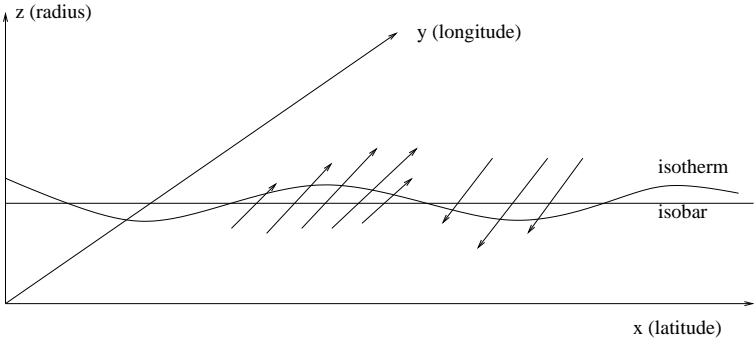}}
\caption{Thermal wind in the cartesian analog of rotating radiative
envelope.}
\label{setup}
\end{figure}

The configuration is shown in figure \ref{setup}; to keep things as simple as
possible Busse considers an axisymmetric flow whose cartesian analog is:

\[ \vv=\la v(x,z)\ey + \rot(\Psi(x,z)\ey)\ra e^{\lambda t}\]
where $\lambda=-\sigma+i\omega$; considering the flow is of small
amplitude all nonlinear terms are neglected; the equation are further
simplified by using the Boussinesq approximation. Although the system
seems to be much simpler than real stars, the heart of the problem, as
discovered by von Zeipel, is conserved. Baroclinicity, as it is called,
is still there. Casting the equations of motions into a single scalar
equation yields:

\[ \lc (\omega^2+2i\omega\sigma)\Delta - 4\Omega^2\ddz{}
-N^2\ddx{}\lp1-\frac{i\kappa}{\omega}\Delta\rp\rc\Psi = 0 \]
which can be solved by Fourier expansions in the horizontal direction.
Solutions are split into different sets of damped modes. The main ones
are the gravito-inertial modes:

\beqa
\Psi &=& \cos(mkx)\sin\lp n\pi(z/d+1/2)\rp \\
V &=& \frac{2\Omega n \pi}{i\omega}\cos(mkx)\cos\lp n\pi(z/d+1/2)\rp \\
\theta &=& \frac{\beta m k}{i\omega\rho_0}\sin(mkx)\sin\lp
n\pi(z/d+1/2)\rp \\
\omega^2 &=& \frac{4\Omega^2n^2+N^2(mkd/\pi)}{n^2+(kd/\pi)^2} \\
\sigma &=& \kappa k^2m^2\frac{N^2}{2\omega^2}
\eeqa

and the baroclinic modes:

\beqa
\theta &=& \sin(mkx)\sin\lp n\pi(z/d+1/2)\rp \\
V &=& \frac{mk\rho\alpha gd}{2\Omega k \pi}\cos(mkx)\cos\lp
n\pi(z/d+1/2)\rp \\
\Psi &=& \od{\kappa/d^2}                   \\
\omega^2 &=& 0 \\
\sigma &=&
\frac{\kappa}{d^2}\;\;\frac{n^2\pi^2+m^2k^2}{1+{N^2}/{4\Omega^2}\lp{mk}/{n\pi}\rp^2}
\eeqa
These sets of modes should be completed by the geostrophic modes and pure
thermal modes.

The fact that the Fourier basis is complete implies the completeness of the
set of modes and therefore the possibility of decomposing any
initial condition into a linear combination of the modes. Thus the
time scale of the transient is controled by the least-damped modes. To
obtain the corresponding spherical analog of this model we should make
the correspondance

\[ d\sim R, \qquad k\sim 2/R \]

Two damping rates associated with the two (main) sets of modes appear

\begin{equation}
\sigma_{GI} \sim \frac{4\kappa}{R^2}, \qquad \sigma_{\rm baro} \sim
\frac{(\pi^2+4)\pi^2}{4}\eta\frac{\kappa}{R^2} \label{eqeta}\end{equation}
where $\eta = \frac{4\Omega^2}{N^2} < 1$. Thus we find two time scales:

\begin{itemize}
\item $T_{KH} = \frac{R^2}{\kappa}\equiv$ The thermal diffusion time or
Kelvin-Helmoltz time scale
\item $T_{ED} = T_{KH}/\eta\equiv$ The Eddington-Sweet time scale.
\end{itemize}

For rapid rotators $\eta\sim 1$ and all transient are damped on the
Kelvin-Helmoltz time, while for slow rotators $\eta\ll1$ baroclinic
modes decay very slowly and initial conditions play a crucial role.

What about meridian circulation ? In this model it appears through 
$\Psi$. We see that there is some amplitude associated with
gravito-inertial modes, but this is oscillatory motion which disappears
after some diffusion time. Rather, the decay of baroclinic modes,
which are characterized by no oscillation may be considered closer to
what was imagined by Eddington, Vogt and Sweet. However, we see that
either the star is rapidly rotating and such flows disappears on a
Kelvin-Helmoltz time or it is slowly rotating and it keeps a strong
memory of the initial conditions.

\begin{figure}
\centerline{\includegraphics[width=10cm,angle=0]{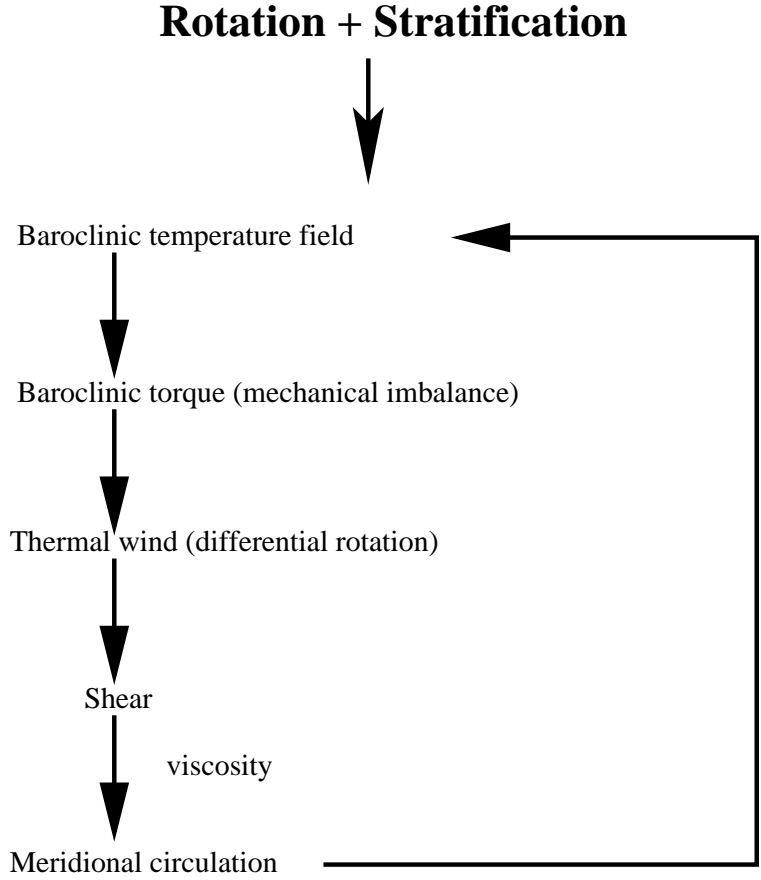}}
\caption{Busse scenario; see comments in fig.~\ref{Schem_zahn}.}
\label{Schem_buss}
\end{figure}

Finally, to make the discussion complete, Busse considered the Ekman
layers which, through Ekman (viscous) pumping, can induce some
circulation, but considering the small value of viscosity, such flows
turn out to be even weaker. His paper ends with a discussion of the
stability of the steady flow (the thermal wind), a stability which
depends on the amplitude of course.
The scenario proposed by Busse may be illustrated as in figure
\ref{Schem_buss}.

Back to the title of Busse's paper ``Do Eddington-Sweet circulations
exist?" we clearly see that the answer is negative. In
a following paper \cite[]{busse82} aimed at popularizing this idea,
Busse hammers it in: ``{\em ... there are no special limits in which
the Eddington-Sweet theory provides a correct solution of the basic
equations}''. This strong conclusion comes after forty years (since the work
of Randers) during which the main stream literature has been forgetting
basic physics principles !

In fact, the work of Sweet correctly derived the now well-known
'Eddington-Sweet' time scale, which is the time it takes for the longest
transient to be damped out. After \cite{sweet50}, this time scale is
also understood as the time it takes for a fluid element to cross a
distance like the radius of the star; we shall see below (\S\ref{PG})
that this can be true only in an unrealistic case.

The work of Busse however considers an idealized situation where the
transport coefficient are of molecular or radiative origin. Hence, the
Prandtl number is very small. Actually, even a radiative zone may be the seat
of some turbulence generated by (non-axisymmetric) baroclinic
instabilities. Indeed, the thermal wind is basically a shear flow prone
to shear instabilities. We thus expect that some turbulence takes place
and enhances  strongly the diffusion of momentum in stellar plasma.

But, as we pointed out, the presence of viscosity (or any local diffusive
transport) changes much the situation since it permits the existence of a
steady meridian circulation. We shall now examine this question
through the work of \cite{zahn92}.

\subsection{Zahn 1992}

The work of \cite{zahn92} is basically the implementation of Busse's
approach in the framework of spherically symmetric stars but taking into
account turbulent viscosity.

The main idea is to assume that, thanks to the stable radial
stratification of a radiative zone, turbulence is essentially horizontal.
Hence, turbulent diffusion is strongly anisotropic, being most vigorous in
the horizontal direction, implying a  differential rotation developed
preferentially in the radial direction.

The model concentrates on three equations:

\begin{itemize}
\item The advection-diffusion balance of the angular momentum flux:

\[\frac{1}{5r^2}\dr{}(\rho r^4 U \Omega) + \frac{1}{r^2}\dr{}\lp r^4
\nu_h\dr{\Omega}\rp = 0 \]
where $v_r = U(r)P_2(\cth)$; this form of the velocity field retains only
the first term in a Legendre polynomial expansion.

\item The balance of torques, source of the thermal wind,

\[ -\frac{\na\rho\wedge\na P}{\rho^2} = \na\Omega^2\wedge s\es\]

\item The advection-diffusion balance of the thermal flux

\beq \Div(\khi\na T) +\rho\eps = \rho c_v \vv\cdot\na T\eeqn{energy_eq}
\end{itemize}
These equations are of course completed by the equation of mass
conservation.

Zahn then imposes an important massage to the equation of energy
\eq{energy_eq} in order to obtain the expression of the thermal imbalance
$\Div(\khi\na T) +\rho\eps$ in terms of its barotropic and baroclinic
components. Each of these components are, like the velocity field,
expanded onto Legendre polynomials and only the $P_2(\cth)$-component is
retained; the thermal imbalance is thus written:

\[ \Div(\khi\na T) +\rho\eps =
\overline{\rho}\frac{L}{M}(E_\Omega+E_\mu)P_2(\cth)\]
where $E_\Omega$ and $E_\mu$ are non-dimensional functions of rotation
and compositional gradient
respectively. 

In section \S3.5 of the paper Zahn discusses the case of the meridian
velocity since its expression is easily derived from \eq{energy_eq}.
There, he shows how to recover the expressions formely derived by
\cite{gratton45, sweet50, opik51, mestel66} when the rotation is assumed
uniform. Of course such a rotation cannot exist by itself and need to be
imposed by some trick (a magnetic field for instance although in such a
case even the circulation might be suppressed, see a discussion of the
case of magnetic stars in \cite{mestel03}). However, as shown by Busse,
if the trick is removed, these expression of the meridian circulation
are pure non-sense. Indeed, a state of rigid rotation is never reach and
terms coming from the $\Omega$-gradients will just cancel the $E_\Omega$
term ! In a steady state, radial velocity vanishes if there is no
viscosity. The casual reader should therefore not overlook section \S4.1
of Zahn's paper where the dependence between circulation, viscosity and
differential rotation is given explicitly (see equation 4.6).
This point is emphasized in \cite{MZ98}: ``{\em ... not only $E_\mu$ but
also $E_\Omega$ evolves to stop the circulation in a star which does not
lose angular momentum}".

\begin{figure}
\centerline{\includegraphics[width=10cm,angle=0]{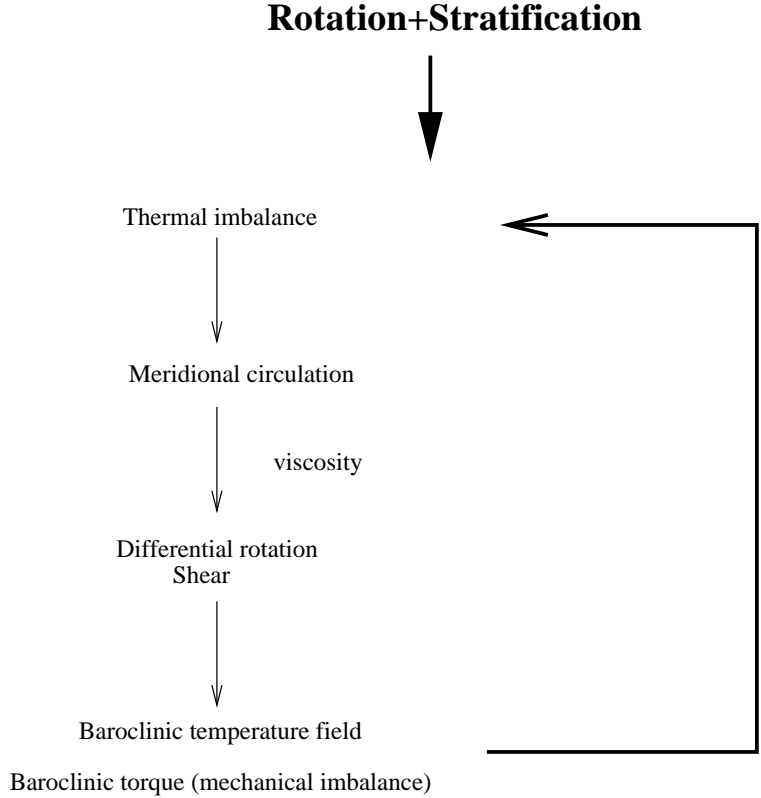}}
\caption{Zahn's approach. This is the same as Busse's one but put in
another ordre. Each arrow symbolises an equation; therefore their
orientation has no meaning and could be reversed! however, it is often
seen in the literature that thermal imbalance implies meridian
circulation. Such a statement is pure non-sense if isolated from the
others; we have seen that it would be much better to say that
``differential rotation implies, through viscosity, meridian
circulation". In fact, this diagram just shows that we face a system of
coupled partial differential equations.}
\label{Schem_zahn}
\end{figure}

\subsection{A further example}

In order to further illustrate the relations between circulation,
viscosity and differential rotation, we shall focus on another simple
model, in the style of Busse's but which retains viscosity and spherical
geometry. 

The set-up is that of a rotating, self-gravitating, full fluid sphere
of nearly constant density so as to be allowed to use the Boussinesq
approximation. The boundaries of the domain remain spherical whatever the
rotation speed.

Within such a set-up, fluid motion, in the rotating frame, is governed
by the following equations:

\begin{eqnarray*}
2\vO\wedge\vv + \vv\cdot\na\vv &=& -\frac{1}{\rho}\na \delta P +
\frac{\delta\rho}{\rho}(\vg+\Omega^2s\es)+\nu\Delta\vv \\
\vv\cdot\na \delta T &=& \kappa\Delta \delta T+Q \\
\Div\vv &=& 0\\
\frac{\delta\rho}{\rho} &=& -\alpha\delta T
\end{eqnarray*}
where $\vv$ is the velocity field, $\vO=\Omega \ez$ the angular rotation
of the frame, $\vg$ the local gravity, $\nu$ the kinematic viscosity,
$\kappa$ the thermal diffusivity, $\alpha$ the thermal dilation
coefficient.

Using $R$, the radius of the sphere, as the length scale, $(2\Omega)^{-1}$
as the time scale, $T_*$ as the temperature scale, the non-dimensional
form of the equation is:

\greq
\ez\wedge\vu + \vu\cdot\na\vu = -\na p -\RA\Theta(-r\er+\eps s\es) +
E\Delta\vu \\
\vu\cdot\na\Theta = E_T\Delta \Theta + q\\
\Div\vu=0
\egreqn{base1}
with the non-dimensional numbers

\[ \RA=\frac{\alpha T_* g}{4\Omega^2R},
\qquad \eps = \frac{\Omega^2R}{g}\]
which control the driving and the other numbers

\[E = \frac{\nu}{2\Omega R^2}, \qquad E_T=\frac{\kappa}{2\Omega R^2},
\]
which control the diffusion.

The system \eq{base1} is simplified by elimination of the barotropic
component of the background which is spherically symmetric and in
hydrostatic equilibrium; we thus write:

\begin{eqnarray*}
\vu=\eps\vu_1, \qquad \Theta=\Theta_0(r)+\eps\Theta_1, \qquad
p=p_0(r)+\eps p_1
\end{eqnarray*}
Furthermore, if we introduce the \BV\ frequency

\[ N^2_T = \alpha\dnr{T_0}g = 4\Omega^2 \RA r\Theta'_0(r)\]
and set $\RA\, r\Theta'_0= rn^2(r)$, the equation of motion may be cast
into:

\greq
\rot\lc\ez\wedge\vu+\eps\vu\cdot\na\vu - 
(r\er-\eps s\es)\Theta_1 -E\Delta\vu\rc
 = -rn^2\sth\cth\ephi \\
\\
n^2u_r + \eps\vu\cdot\na\Theta_1 = E_T\Delta \Theta_1\\
\\
\Div\vu=0
\egreq
The equation of vorticity clearly shows that the motion is driven by the
baroclinic torque, $-rn^2\sth\cth\ephi$ issued from the set-up. We are
therefore facing the problem of a steady forced flow.
We may further simplify the problem by
considering small $\eps$, i.e. small centrifugal effect; at first order
the flow is controlled by the linear equations:

\greq
\rot\lp\ez\wedge\vu - r\er\Theta_1 -E\Delta\vu\rp =  -rn^2\sth\cth\ephi \\
\\
n^2u_r  = E_T\Delta \Theta_1\\
\\
\Div\vu=0
\egreqn{sys_lin}
It is quite easy to see that if viscosity is neglected, the thermal wind

\greq \vu = \lp -s\int n^2(r) dr + F(s)\rp\ephi,\\
\\
 \Theta_1=0
\egreqn{vt0}
is a solution of the equations. Here $s=r\sin\theta$ is the radial
cylindrical coordinate and $F(s)$ an
arbitrary function describing a pure geostrophic solution. We thus find
again that in such a baroclinic steady flow, no meridian circulation is
allowed if viscosity is zero.

The dependance between viscosity and meridian circulation may be made
even more explicit if we project the system \eq{sys_lin} on the spherical
harmonics and have a look to first terms.  Following \cite{rieu87}, we set

\[\vu=\sum_{l=0}^{+\infty}\sum_{m=-l}^{+l} \ulm\RL+\vlm\SL+\wlm\TL,
\qquad \Theta_1=\sum_{l=0}^{+\infty}\sum_{m=-l}^{+l} \theta^\ell_m\YL \]
where

\[\RL=\YL\vec{e}_{r},\qquad \SL=\na\YL,\qquad \TL=\na\times\RL \]
with $\YL$ being the usual normalized spherical harmonic function. The
projection of the equation leads to 

\[\left\{ \begin{array}{l} 
E\Delta _lw^l_m+A^l_{l-1}r^{l-1}\drr\biggl(
\frac{u_m^{l-1}}{r^{l-2}}\biggr)
+A^l_{l+1}r^{-l-2}\drr\biggl( r^{l+3}u_m^{l+1}\biggr) =0  \\
\\ 
E\Delta _l\Delta _l(ru^l_m)= B^l_{l-1}r^{l-1}\drr
\biggl(\frac{w_m^{l-1}}{r^{l-1}}\biggr)
+ B^l_{l+1}r^{-l-2} \drr\biggl( r^{l+2}w_m^{l+1}\biggr) \\
\qqqquad\qqqquad +l(l+1)\theta^l_m - 6rn^2(r) N_2\delta_{l2} \\
\\
E_T\Delta_\ell\theta^l_m  -n^2(r)u^l_m =0 \\
\end{array} \right. \]
where $N_2=\frac{1}{6}\sqrt{\frac{16\pi}{5}}$ and $A^l_{l-1}, B^l_{l-1},
...$ are coupling coefficients. This is the set of
ordinary differential equations which can be solved numerically to obtain
the full solution. However, for the pedagogical matters we are interested
in, it is sufficient to consider $\ell\leq2$. We thus have:

\[\left\{ \begin{array}{l} 
E\Delta_1 w^1 + A^1_2r^{-3}\drr\biggl( r^{4}u^{2}\biggr) =0  \\
\\ 
E\Delta_2\Delta_2(ru^2)=
 B^2_{1}r^{1}\drr
\biggl(\frac{w^{1}}{r}\biggr)+6\theta^2 - 6rn^2 N_2\\
\\
E_T\Delta_2\theta^2  -n^2u^2 =0 \\
\end{array} \right. \]
In the axisymmetric problem which is considered here $u^\ell$'s
represent the radial velocity and therefore meridian velocity while
$w^\ell$'s describe the azimuthal velocity.
The first equation shows that if viscosity is zero, i.e. $E=0$,
then $u^2=0$, hence no meridian flow occurs and the temperature
field is in equilibrium; the gradient of the azimuthal flow,
$B^2_{1}r\drr\biggl(\frac{w^{1}}{r}\biggr)$ just compensates
the baroclinic torque $- rn^2 6N_2$.

From these equations we may see that, when viscosity is taken into
account, the shear drives a meridian circulation which in turns drives
thermal imbalance !

\subsection{Some additional remarks}\label{PG}

The foregoing example may be completed by the work of \cite{G02} which
presents numerical solutions of the same problem but in a more realistic
set-up: there, the Boussinesq approximation is relaxed; instead the
anelastic approximation is used to  filter out sound waves and permits
the existence strong density contrasts of the background. Thus, radial profiles
of pressure, temperature or density, are more stellar or solar like.

\cite{G02} shows that steady flows are controlled by the non-dimensional
ratio $\lambda={\cal P}N^2/\Omega^2= {\cal P}/\eta$ (see \eq{eqeta} for
the definition of $\eta$). For $\lambda\gg1$,
i.e. for slow rotation, it is shown that the scaling of meridian velocity
is indeed that of Eddington-Sweet, this circulation is confined to a layer
close to the outer boundary. In the case of a fast rotation $\lambda\ll1$
however, Garaud finds that meridian flows scale as $\nu/r$ in agreement
with our foregoing discussion.

Could it be that Sweet's result applies to slowly rotating stars? In
this case indeed the result found by Garaud gives an amplitude of the
circulation which is independent of viscosity and which corresponds
to a turn over time equal to the Eddington-Sweet time. Let us first
note that extremely slow rotation are needed; indeed, $\lambda\gg1$
means $\eta \ll{\cal P}$, while we know that ${\cal P}\ll 1$ in a
radiative zone. Hence, this result applies only when Eddington-Sweet
time is at least ${\cal P}^{-1}$ greater than thermal diffusion time
(i.e. Kelvin-Helmoltz time); in such conditions, it is clear that
Eddington-Sweet time is much longer than the actual lifetime of the
star. Therefore, back to Busse's result, it is also clear that such a
steady state is never reached.

\subsection{Unsteady situations}

This discussion brings us to the question of the role of initial
conditions. Until now we focused our attention to the case of steady
flows, however if the star is initially rotating slowly so that
Eddington-Sweet time exceeds nuclear time, it is clear that, according
to Busse, the rotational state of the star will be marked by the initial
conditions. In such case modeling the actual dynamical state of star
requires the knowledge of an enormous number of parameters, namely the
initial velocity field (at least!).

However, as shown by observations, young stars are rather rapid rotators
and therefore we may expect that they reach a sort of universal state of
differential rotation. But it is also well known that in the course of
their lives, stars lose angular momentum and thus progressively slow
down their rotation.

\subsection{Spin-down processes}

This remark brings us to the role of winds and angular momentum losses.
As already stressed by \cite{zahn92}, this is
certainly the main source of mixing for main sequence stars. The
supposed universal differential rotation which comes out of the balance
between baroclinic torque and (turbulent) viscosity is thus perturbed by
the shear produced by the angular momentum loss. In fact this becomes
the dominant source of turbulence in radiative zone and the dynamical
state of the star is now controlled by the mechanisms which insure the
angular momentum transfer between the different layers. It may be
turbulent diffusion, but also gravity waves emitted by a neighbouring
convective zone \cite[see][]{TC03}.

To give an intuitive idea of how a spin down process works, it is useful
to go back to the laboratory. Consider a fluid contained
in some rotating container; at some time when the fluid is in perfect
solid rotation, the rotation of the container is slowed down by some
small amount (so that all the spin-down process remains controlled by
linear equations). After some rotation, Ekman boundary layers form on
the wall of the container which are not parallel to the rotation axis.
These layers are the seat of shear flow parallel to the wall and when
the wall imposes a no-slip boundary conditions, dissipation is very
intense there. However, in addition to this dissipative process, Ekman
layers also pump the fluid into or out of the layer. By this process, the
layer drives a circulation in the bulk of the fluid mass, thus inducing a
transport of angular momentum from the bulk to the layer. In a
laboratory experiment the friction of the fluid, through viscosity, on
the container's wall, finally removes the angular momentum from the fluid.
The circulation, induced by the Ekman layer, plays a fundamental role in
the spin-down process since it governs the time-scale of the process.
This time scale is indeed

\[(2\Omega)^{-1}\lp\frac{2\Omega L^2}{\nu}\rp^{1/2} \]
much smaller than the diffusion time scale $L^2/\nu$.

Back to the case of stars there are some differences, which still need
to be investigated in detail, but the gross features are certainly quite
similar.

There are of course no rigid wall at the surface of a star; however, if
we consider the case of a solar type star with an outer envelope
threaded by magnetic fields, it is likely that the convective envelope
will lose quite rapidly its angular momentum and behave like a rigid
body stressing the lower radiative zone. There everything depends on the
viscosity : if strong enough it can extract angular momentum at the
desired rate imposed by the wind. On the contrary, the shear layers may
become unstable and turbulence should develop enhancing the viscosity;
the foregoing situation may be restored and extraction takes the imposed
rate; however, shear has a maximum value and therefore the stress (equal
to shear times viscosity) may not be high enough; in such a case, called
strong wind in \cite{zahn92}, the radiative zone will slow down at a
rate controlled by turbulent diffusive processes.

In this description we have not described the effect of stratification
and heat diffusion which make the whole picture a little more complex.
In fact the whole problem has been only touched upon and the whole
non-linear scenario is waiting exploration. Basic references on spin-down
process in the laboratory context are \cite{Green69} or
\cite{rieutord97}; a discussion on the effect of stratification and
small Prandtl number can be found in \cite{fried76}; an
application to the stellar (solar) context has been tempted by
\cite{ZTM97}.

\section{Summarizing essential points}

To end this lecture I would like to stress the most important points,
those one should keep in mind when dealing with rotating radiative zones.

First we need to stress that

\[ \rho c_v\vv\cdot\na T =\Div(\khi\na T) +\varepsilon \]
is not an equation determining the velocity field. It determines the
temperature once $\vv$ is given by the dynamics; there is no other mean
to determine the temperature. Giving an a-priori-approximate-solution for
$T$ and using it to derive $\vv$ is just an incorrect reasoning. Any
approximate solution of $\vv$ needs to be a solution of both the
momentum and energy equation (or a combination of both).

It is another error to specify to rotation state (usually assumed rigid)
as it will quickly (on a Kelvin-Helmoltz time however) evolve to a
differential rotation state thanks to the diffusion of heat. Even if
rotation is slow, such a reasoning is misleading since the rotation
state at time $t$ is the result of an evolution which cannot be
discarded; initial conditions, which are usually a fast and differential
rotation are not forgotten by a presently slowly rotating star.

Finally, we have shown that viscosity (or any source of diffusion of
momentum) plays a crucial part in the dynamical evolution of a star, in
particular in mixing processes. An isolated star (no wind) cannot
sustain a large-scale circulation in a quasi-steady state if no
diffusion process occurs.

Finally, let us stress that the loss of angular momentum is another
crucial aspect for the mixing processes since these losses also drive
meridian circulation. It underlines again the importance of the
dynamical history of a star when one tries to understand its present
surface abundances.

\begin{acknowledgements}
I would like to thank Jean-Paul Zahn for the many interesting discussions we
had on this subject and his kind reading of this manuscript.
\end{acknowledgements}

\bibliography{biblio}

\end{document}